\documentclass[prl,twocolumn, aps,amssymb,showpacs]{revtex4-1}

\usepackage{graphicx}
\usepackage{amsmath,amssymb}
\usepackage{cancel}
\usepackage{color}
\usepackage{pifont}
\usepackage{float}

\newcommand{\ket}[1]{\ensuremath{\left| #1 \right\rangle}}

\usepackage{ulem}
\usepackage{cancel,ifthen}
\newcommand{\cmnt}[2][NoInPuT]{\ifthenelse{\equal{#1}{NoInPuT}}{}{{\color{red}\sout{#1}}} {\color{blue} #2}}

\usepackage{hyperref}

\begin{document}

\title{Interferometric Approach to Probing Fast Scrambling}
\author{N. Y. Yao$^{1}$, F. Grusdt$^{2}$, B. Swingle$^{3}$, M. D. Lukin$^{2}$, D. M. Stamper-Kurn$^{1}$, J. E. Moore$^{1}$, E. Demler$^{2}$}
\affiliation{$^{1}$Department of Physics, University of California Berkeley, Berkeley, CA 94720, U.S.A.}
\affiliation{$^{2}$Department of Physics, Harvard University, Cambridge, MA 02138, U.S.A.}
\affiliation{$^{3}$Stanford Institute for Theoretical Physics and Department of Physics, Stanford University, Stanford, California 94305, U.S.A.}

\begin{abstract}
Out-of-time-order correlation functions provide a proxy for diagnosing chaos in quantum systems. We propose and analyze  an interferometric scheme for their measurement, using only local quantum control and no reverse time evolution. Our approach utilizes a combination of Ramsey interferometry and the recently demonstrated ability to directly measure Renyi entropies.
To implement our scheme, we present a pair of cold-atom-based experimental blueprints; moreover, we demonstrate that within these systems, one can naturally realize the transverse-field Sherrington-Kirkpatrick (TFSK) model, which exhibits certain similarities  with fast scrambling black holes.
We perform a detailed numerical study of scrambling in the TFSK model, observing an interesting interplay between the fast scrambling bound and the onset of spin-glass order.

\end{abstract}


\maketitle

Much of statistical mechanics rests on the assumption that generic systems will, after sufficient time, arrive at a state close to thermal equilibrium.
In isolated quantum systems, the approach to equilibrium is characterized by the spreading of entanglement and quantum information.
That there might exist fundamental limits on the rate of thermalization has a long history \cite{wiener1938homogeneous,hoover1985canonical,eckmann1985ergodic,Anderson:1958p7531,srednicki1994chaos,rigol2008thermalization}.
At one extreme are strongly disordered systems, where thermalization is absent and quantum information spreads slowly \cite{Basko:2006hh,Gornyi:2005fv,oganesyan07,palhuse10,Schreiber:2015aa,kondov2015disorder,couplingBordia16,Choi1547}.
At the other extreme, certain gauge theories appear to spread quantum information very rapidly \cite{Sekino:2008he,Shenker:2013pqa,Maldacena2015}.
However, these gauge theories are  special---their thermal states are ``holographically dual" to black holes in Einstein gravity \cite{Maldacena1997zz}. They are also highly symmetrical, display scale invariant physics, and do not order at low temperatures despite strong interactions.
%

Thus, a key question is: Where do \emph{typical} interacting systems fall between these two extremes?
The lack of general theoretical tools in this context suggests that experiments will be essential to explore the nature of information spreading or ``scrambling'' in many-body systems \cite{Sekino:2008he}.
More precisely, scrambling describes the delocalization of quantum information over all of a system's degrees of freedom.
The analog of scrambling in a classical system is chaos and is diagnosed by the butterfly effect, which describes the exponential sensitivity of a particle's motion to small changes in its initial conditions.
As an example, the Poisson bracket of position and momentum, $\{x(t),p\} = \partial x(t) / \partial x(0)$, can be used to quantify the butterfly effect \cite{wiener1938homogeneous}.
The strength of quantum scrambling can  similarly be diagnosed using the commutator, $C(t) = \langle [W(t),V(0)]^\dagger [W(t),V(0)] \rangle$, where $V$ and $W$ are unitary operators \cite{larkin1969quasiclassical, Sekino:2008he, Shenker:2013pqa, almheiri2013apologia, kitaev15}.

The functional onset of scrambling is particularly intriguing, with certain systems exhibiting a parametric period of exponential growth, $C(t) \sim e^{\lambda_L t}$ \cite{sachdev1993gapless, kitaev15, maldacena2016comments, polchinski2016spectrum}.
In semi-classical systems, $\lambda_L$ can be interpreted as a Lyapunov exponent characterizing the strength of chaos.
For gauge theories dual to black holes, there exists a parameter $N$, such that the large-$N$ limit is  semi-classical and $\lambda_L = 2\pi k_B T/ \hbar$, saturating a recently proposed upper bound on the rate of scrambling \cite{Maldacena2015}.

\begin{figure}
\includegraphics[width=2.5in]{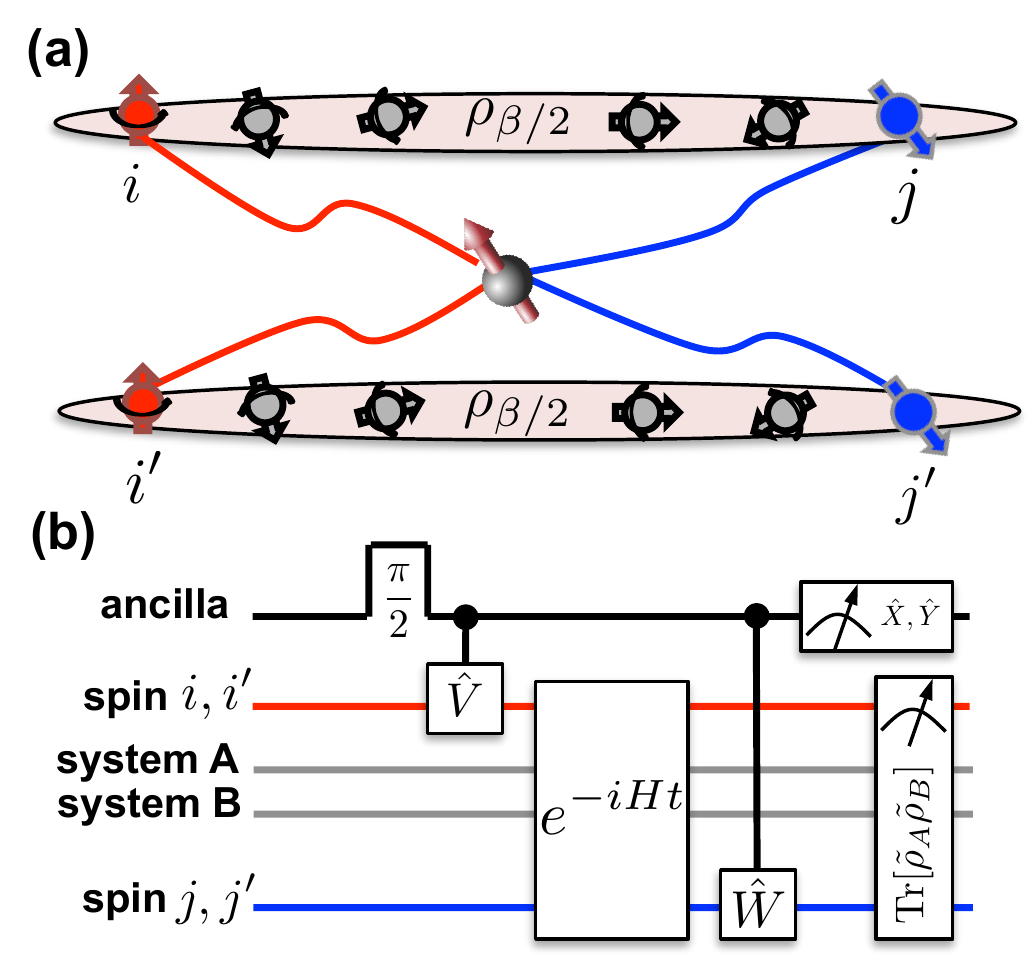}
\caption{%
\label{fig:Intro}
(a) Schematic of our interferometric protocol. An ancilla qubit is coupled to two copies of a quantum system (prepared at inverse temperature $\beta/2$) at different locations $i,j$ and $i',j'$. By performing ancilla-state dependent interactions both before and after time evolution, one can measure out-of-time-order correlation functions.
(b) Circuit diagram illustrating the interferometric protocol. After preparing the ancilla in $1/\sqrt{2}(\ket{\uparrow}+ \ket{\downarrow}$), controlled unitaries are performed, first conditioned on $\ket{\uparrow}$ ($V$) and then conditioned on $\ket{\downarrow}$ ($W$). Simultaneous measurement of $\sigma_x$ on the ancilla and SWAP on the systems results in the correlator $F(t)$. }
\end{figure}

While intriguing, the scrambling behavior of systems dual to black holes is only one extreme of a largely unexplored landscape.
For example, although temperature provides the only natural energy scale for a black hole, generic quantum many-body systems can exhibit a multitude of additional energy scales governed, e.g. by microscopic couplings and/or low temperature order. Whether the presence of these additional scales precludes fast scrambling, at or near the conjectured bound, remains an open question.
Experimental exploration is thus crucial to search for new ``universality classes" of scrambling behavior and to develop a comprehensive picture of information scrambling in systems realized in nature.

To directly measure the commutator $C(t)$ amounts to a measurement of an out-of-time-order correlation function, $F(t) = \langle W^\dagger(t) V^\dagger(0) W(t) V(0) \rangle$ where $C = 2 - 2 \,\text{Re}[F]$.
Experimentally, the measurement of $F(t)$ is extremely challenging because it effectively involves reversing the flow of time at various points in the measurement, or equivalently, time-evolving with both the Hamiltonian, $H$, and its negative counterpart, $-H$.
A protocol using both forward and backward time evolution has been developed to measure $F$ \cite{Swingle2016}; while the necessary ingredients can be engineered in a variety of quantum optical systems, a complementary approach not involving backwards time evolution would be advantageous.

%

In this Letter, we introduce an interferometric approach to measure out-of-time-order correlation functions which requires only local quantum control and \emph{no} reverse time evolution (Fig.~1a).
Our method utilizes a combination of Ramsey interferometry and the recently demonstrated experimental capability to measure Renyi entropies \cite{Daley2012,Abanin2012Ent,Islam2015}.
Besides not requiring reverse time evolution, the scheme also enables the measurement of so-called thermally regulated out-of-time-order correlations, e.g. $\text{Tr}[ \sqrt{\rho} W^\dagger(t) V^\dagger(0) \sqrt{\rho} W(t) V(0) ]$ with $\rho$ the thermal state.
This regulated correlator  is believed to exhibit identical early time scrambling signatures and helps to enable analytical calculations in certain limits \cite{Maldacena2015, kitaev15, maldacena2016comments, polchinski2016spectrum}.
Because our scheme requires the measurement of a many-body overlap, it is naturally suited to study systems with relatively low entropy.

To this end, we perform a numerical study of scrambling dynamics in the transverse-field Sherrington-Kirkpatrick (TFSK) model  with $N\sim 20$ spins \cite{Sherrington1975,ishii1985effect,fedorov1986quantum,Ray1989},
\begin{equation} \label{tfsk}
H_{\textrm{TFSK}}= - \frac{1}{2} \sum_{i \neq j} J_{ij} \sigma_i^z \sigma_j^z - \Gamma \sum_j \sigma_j^x
\end{equation}
where $\vec{\sigma}$ are Pauli operators. The model consists of $N$ spins in a transverse field $\Gamma$, with random all-to-all Ising interactions, $J_{ij}$, drawn from a normal distribution of zero mean and variance $\sigma = \sqrt{J^2/N}$ \cite{Ray1989}.
The all-to-all couplings mimic a similar aspect of black hole models and open the possibility of observing fast scrambling \cite{Sekino:2008he,sachdev1993gapless, kitaev15, maldacena2016comments, polchinski2016spectrum}.
The numerics depict a rich variety of scrambling behaviors even for relatively small system sizes that should be readily accessible with our  protocol. Intriguingly, we also find a suggestive interplay between the bound on scrambling and the onset of spin-glass order.
Finally, we provide a pair of cold-atom-based experimental blueprints for implementing the TFSK model and measuring the scrambling correlation function.

\emph{General Strategy}---Let us begin by considering a many-body quantum system with Hamiltonian $H$.
The correlator $F(t)$ can be written as $\text{Tr}[ \rho_\beta e^{i H t} W^\dagger(0) e^{-i H t} V^\dagger(0) e^{i H t} W(0) e^{-i H t} V(0)]$, where $\rho_\beta$ is the thermal density matrix of the system and $\beta = 1/k_B T$ is the inverse temperature. 
Noting that $\sqrt{\rho} \sim \rho_{\beta/2}$, our strategy will be to measure
\begin{align}
	\label{eq:hamgeneralf2}
	F_2(t) =  \text{Tr}[  e^{-i H t} V^{\dagger} \rho_{\beta/2} e^{i H t} W^{\dagger} e^{-i H t} V \rho_{\beta/2} e^{i H t} W],
\end{align}
which is directly proportional to the thermally regulated correlator~\footnote{Note that in principle $F(t)$ can also be measured directly using our protocol. In this case, one has to replace the second copy at $\rho_{\beta/2}$ by an infinite temperature system.}.
 Two remarks are in order. First, one expects that the magnitude of the experimental signal associated with $F_2(t)$ will decrease with system size as it corresponds to a many-body overlap. Second, $F_2(t)$ factorizes as $\beta \rightarrow \infty$, implying that it will serve as an accurate proxy for $F(t)$ only at finite energy densities \cite{Maldacena2015}. As will be shown in the numerics, for mesoscopic spin ensembles, neither of the above becomes a limiting concern for intermediate temperatures of interest.

Crucially, the functional form of $F_2(t)$ is suggestive of a 2$^\textrm{nd}$ Renyi entropy measurement~\footnote{If $W=V=I$, then $F_2$ reduces to the 2$^\textrm{nd}$ Renyi entropy of the thermal state at $\beta/2$, $F_2 = e^{-S_2(\beta/2)}$.}.
Such a measurement can be realized by preparing two identical copies of a quantum system and performing a subsequent sequence of individual SWAP operations, $S$, between the copies \cite{Daley2012}; indeed, recent experimental progress has harnessed this approach to  explore entanglement-based signatures of thermalization \cite{Islam2015}.

\begin{figure*}
\includegraphics[width=5.0in]{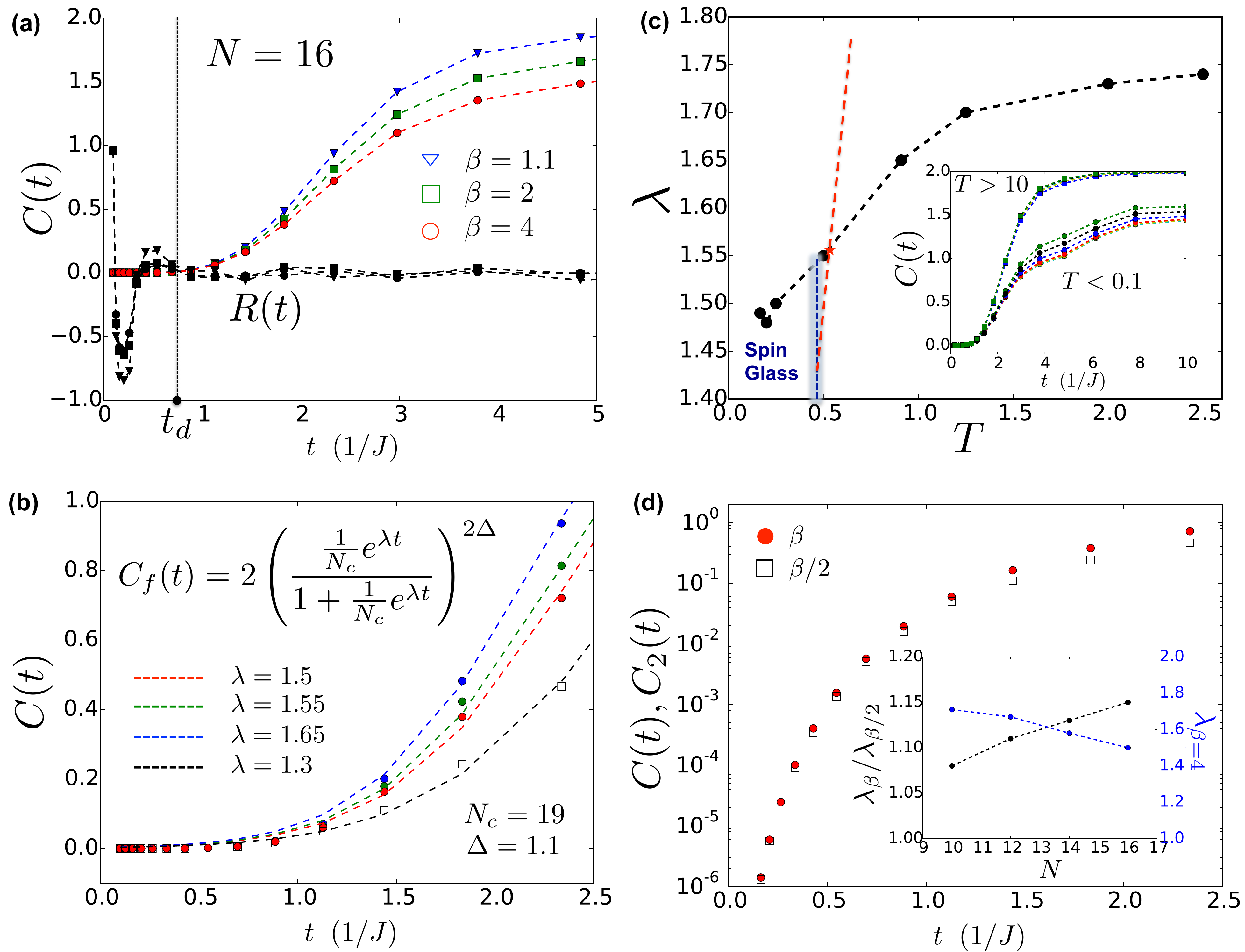}
\caption{%
\label{fig:implementation}
Numerics on the the Sherrington-Kirkpatrick model in a transverse field. For system sizes $N=10,12$, we perform $10^3$ disorder averages, for $N=14$, we perform $10^2$ disorder averages and for $N=16$, we perform $29$ disorder averages. Throughout the simulations, $J=1$ and $\Gamma = 1.35$. (a) Depicts decay of the autocorrelation function $R(t) = \langle \sigma^z_i (t) \sigma^z_i (0) \rangle_\beta$ and growth of the scrambling correlator $C(t) = -\langle [\sigma^z_i(t), \sigma^z_j(0)]^2 \rangle_\beta$ at various inverse temperatures $\beta = 1.1, 2, 4$. The approximate time scale for dissipation is indicated as $t_d$. (b) Numeric fitting to the phenomenological function $C_f(t)$ where $N_c$ and $\Delta$ are fixed for all curves to extract the Lyapunov exponent, $\lambda$, as a function of temperature. (c) Depicts $\lambda$ as a function of temperature. At intermediate temperatures, the growth of $C(t)$ seems to be approximately linear. The temperature of the spin glass transition as determined from Monte Carlo is indicated as the blue dashed line \cite{mukherjee2015classical}. The red dashed line depicts the scrambling bound; since the exponent controlling the early time growth of $C_f(t)$ is $2\Delta \lambda$ (see footnote), we plot $2\pi T / 2\Delta$ to compare directly with the extracted $\lambda$.   The inset shows the differences in $C(t)$ at  low and high temperatures. At low temperatures $C(t)$ does not reach its maximally scrambled value even at late times $\sim 10^2/J$.  (d) Comparison between $C(t)$ and $C_2(t)$ for $\beta = 4$. The inset depicts the ratio of $\lambda$ as extracted from $C(t)$ and $C_2(t)$ as a function of system size (black points) and $\lambda_{\beta=4}$ as a function of system size (blue points). }
\end{figure*}

With this technique in mind, we devise an interferometric approach (Fig.~1b) to measuring out-of-time-order correlation functions by utilizing only a single ancilla qubit (with states $\{ \ket{\downarrow},\ket{\uparrow} \}$) and Hamiltonian evolution.
Let us begin with both copies of the system and the ancilla qubit  prepared in the product state, $\rho(t=0) = (\rho_{\beta/2})^{\otimes2} \otimes | + \rangle \langle + |$, where  $(\rho_{\beta/2})^{\otimes2}=\rho_{\beta/2} \otimes \rho_{\beta/2}$.
Assuming that the ancilla is capable of exerting a local state-dependent interaction on the system, time evolution then yields,
\begin{multline}
	\label{eq:hamgeneral}
	\rho (t) = \frac{1}{2} [ \left | \uparrow  \right \rangle \left \langle \uparrow  \right |  \otimes ( U_{\uparrow}^\dagger \rho_{\beta/2}  U_{\uparrow} )^{\otimes 2} +  \left | \downarrow  \right \rangle \left \langle \downarrow  \right |  \otimes ( U_{\downarrow}^\dagger \rho_{\beta/2}  U_{\downarrow} )^{\otimes 2}  \\
	+ \left | \uparrow  \right \rangle \left \langle \downarrow  \right |  \otimes ( U_{\uparrow}^\dagger \rho_{\beta/2}  U_{\downarrow} )^{\otimes 2} +  \left | \downarrow  \right \rangle \left \langle \uparrow  \right |  \otimes ( U_{\downarrow}^\dagger \rho_{\beta/2}  U_{\uparrow} )^{\otimes 2}  ]
\end{multline}
where $U_{\uparrow}^\dagger = W^\dagger e^{-iHt}$, $U_{\downarrow}^\dagger =  e^{-iHt} V$ and $V$, $W$ are spatially separated at locations $i$ and $j$. Here, $U_{\uparrow(\downarrow)}^\dagger$ are unitary operators that describe the combination of Hamiltonian time evolution and the ancilla-state-dependent interaction (Fig.~1b).

Assuming that $V$ and $W$ are both Hermitian and unitary, the resulting density matrix contains terms of the form
$ ( U_{\downarrow}^\dagger \rho_{\beta/2}  U_{\uparrow} )^{\otimes 2}= e^{-i Ht} V \rho_{\beta/2} e^{iHt} W \otimes  e^{-i Ht} V \rho_{\beta/2} e^{iHt} W$,
precisely matching the desired form of $F_2(t)$ \footnote{For general unitaries, we can realize Eqn.~2 by performing the conjugate controlled-interaction on one system as compared to the other.}. This suggests that one may be able to extract coherences of the time-evolved density matrix and naturally leads us to consider an interferometric approach based upon Ramsey spectroscopy.

In particular, by independently measuring $\sigma_x \otimes S$ and $\sigma_y \otimes S$ (where $\sigma_{x,y}$ act on the ancilla qubit and SWAP $S$ acts on the two copies of the system), one is able to measure the real and imaginary portions of $F_2(t)$. To see this, let us project $\rho(t)$ onto the eigenstates of $\sigma^x$. Noting that $ \text{Tr}[ \rho_A \rho_B] = \text{Tr} [ S \rho_A \otimes \rho_B]$, we obtain
\begin{align}
\text{Tr} \left[ \rho (t) \sigma^x \otimes S \right] &= \text{Tr} \left[ (e^{-iHt} V \rho_{\beta/2}   e^{ iHt} W)^2 \right] + {\rm c.c.} \nonumber \\  &= \textrm{Re}[F_2(t)].
\end{align}
and $\text{Tr} \left[ \rho (t) \sigma^y \otimes S \right] = \text{Im}[F_2(t)]$.
Having demonstrated the ability to directly measure thermally regulated out-of-time-order correlators, we now turn to a numerical study of scrambling dynamics in the TFSK model.
%




\emph{TFSK Numerics}---We perform exact diagonalization and numerical integration of Hamiltonian [Eqn.~(1)] evolution for systems of up to $N = 16$ spins.
%
As previously discussed, while the model features certain similarities with fast scrambling gauge theories \cite{Sekino:2008he,sachdev1993gapless, kitaev15},  it remains an open question whether its dynamics (e.g. growth of $C(t)$) are controlled by the thermal scale or by microscopics; this owes in part to the existence of a low-temperature spin-glass phase for $\Gamma \lesssim 1.5$ \cite{yamamoto1987perturbation,usadel1987quantum,goldschmidt1990ising,miller1993zero,takahashi2007quantum,mukherjee2015classical}.

To begin probing thermalization dynamics in $H_{\textrm{TFSK}}$, we determine the dissipation time, $t_d$, which characterizes the decay of local two-point correlation functions.
In particular, as shown in Fig.~2a (black lines), we compute the disorder averaged autocorrelation function $R(t) = \langle \sigma^z_i (t) \sigma^z_i (0) \rangle_\beta$; $t_d$ is identified as the time at which correlations have dropped to less than $0.05$ and is approximately independent of temperature for $1<\beta < 4$.

At the same temperatures, one finds that $C(t) = -\langle [\sigma^z_i(t), \sigma^z_j(0)]^2 \rangle$ begins to exhibit a sharp rise for $t > t_d$, indicative of the onset of scrambling (Fig.~2a).
Intuition from the black hole problem suggests that the time to reach a maximally scrambled state is of order $\sim\ln N_c$, where $N_c$ is proportional to the entropy.
This suggests fitting  $C(t)$ with a function of the combined variable $e^{\lambda t}/N_c$; indeed, we observe that the early to intermediate time data can approximately be captured using a phenomenological function
\begin{equation}
C_f(t) = 2 \left ( \frac{\frac{1}{N_c} e^{\lambda t}}{1 + \frac{1}{N_c} e^{\lambda t}   } \right )^{2\Delta},
\end{equation}
where $\Delta$ is an operator-dependent fitting parameter (in the context of black holes, $\Delta$ is the scaling dimension of the operator), and $\lambda$ characterizes the Lyapunov exponent.
To quantitatively analyze the data, we perform a nonlinear regression, keeping $N_c$ and $\Delta$ fixed across all temperatures and extract $\lambda$ as a function of $\beta$ as shown in Fig.~2b,c (note that the dotted lines in Fig.~2b correspond to $C_f(t)$ with extracted parameters).

A few remarks are in order. First, at the system sizes studied, it is hard to distinguish between dissipation and the onset of scrambling, although the data are consistent with a small separation in time-scales.
Second, Fig.~2c provides evidence that for $T \lesssim J$, the temperature dependence of $\lambda$~\footnote{The early time growth of $C_f$ is $e^{2 \Delta \lambda t}$, so the growth exponent is $2 \Delta \lambda$. We find that $ 2\Delta \lambda/ T$ at $T=.5$ is about $6.82$. Although this  is greater than the bound of $2\pi$, we emphasize that the bound is only expected to apply  at large-$N$. This is also somewhat consistent with the observed decrease in $\lambda$ as a function of system size.} becomes approximately linear, although the slope is more than an order of magnitude smaller than the conjectured upper bound \cite{Maldacena2015}. At these temperatures, $\lambda$ exhibits a weak decrease as a function of system size (Fig.~2d, inset); intriguingly, this trend is consistent with a recent holographic calculation of $1/N$ corrections to $\lambda_L$ \cite{fitzpatrick2016quantum}.

At even lower temperature, near or below the spin glass transition, $C(t)$ does not reach its maximally scrambled value of $2$  at  late times (Fig.~2c, inset). Moreover, for these finite size systems, there is a near collision between the onset of spin-glass order and the temperature at which the value of $\lambda$ seems to violate the scrambling bound (Fig.~2c, red dashed line). That the value of $\lambda$ below the transition seems to be well above the bound is somewhat surprising as one might have expected scrambling to slow once the spins freeze.

For high temperatures, $T \gtrsim J$, the data show evidence of saturation consistent with scrambling being governed by the microscopic scale.
Finally, Fig~2d depicts a direct comparison between $C(t)$ and $C_2(t) = -\text{Tr}[ \rho_{\beta/2} [W(t),V] \rho_{\beta/2} [W(t),V]]$.  While they agree at short times, differences begin to accumulate at late times.  The ratio of the exponents, $\lambda_\beta$ and $\lambda_{\beta/2}$, extracted from $C(t)$ and $C_2(t)$ respectively (Fig.~2b, Fig.~2d, inset), increases weakly with system size, indicating subtleties in  scaling at larger systems. Crucially, extrapolating the ratio of $C(t)$ and $C_2(t)$ at finite $\mathcal{O}(1)$ times suggests that the signal to noise observed in current generation experiments should allow for exploration of system sizes up to  $N \sim 50$ spins, well beyond the capability of numerics.
%

\emph{Experimental Implementation}---We now propose two experimental blueprints for realizing our interferometric protocol, where the underlying system is composed of an array of ultracold atoms trapped in a 1D optical array (Fig.~3).
First, we will describe an approach based upon Feshbach interactions and second, we will consider a method that utilizes a Rydberg-blockade gate \cite{Jaksch2000,Lukin2001,Urban2009}. Focusing on the latter, we will further demonstrate that this system naturally simulates the TFSK model.

\begin{figure}
\includegraphics[width=2.5in]{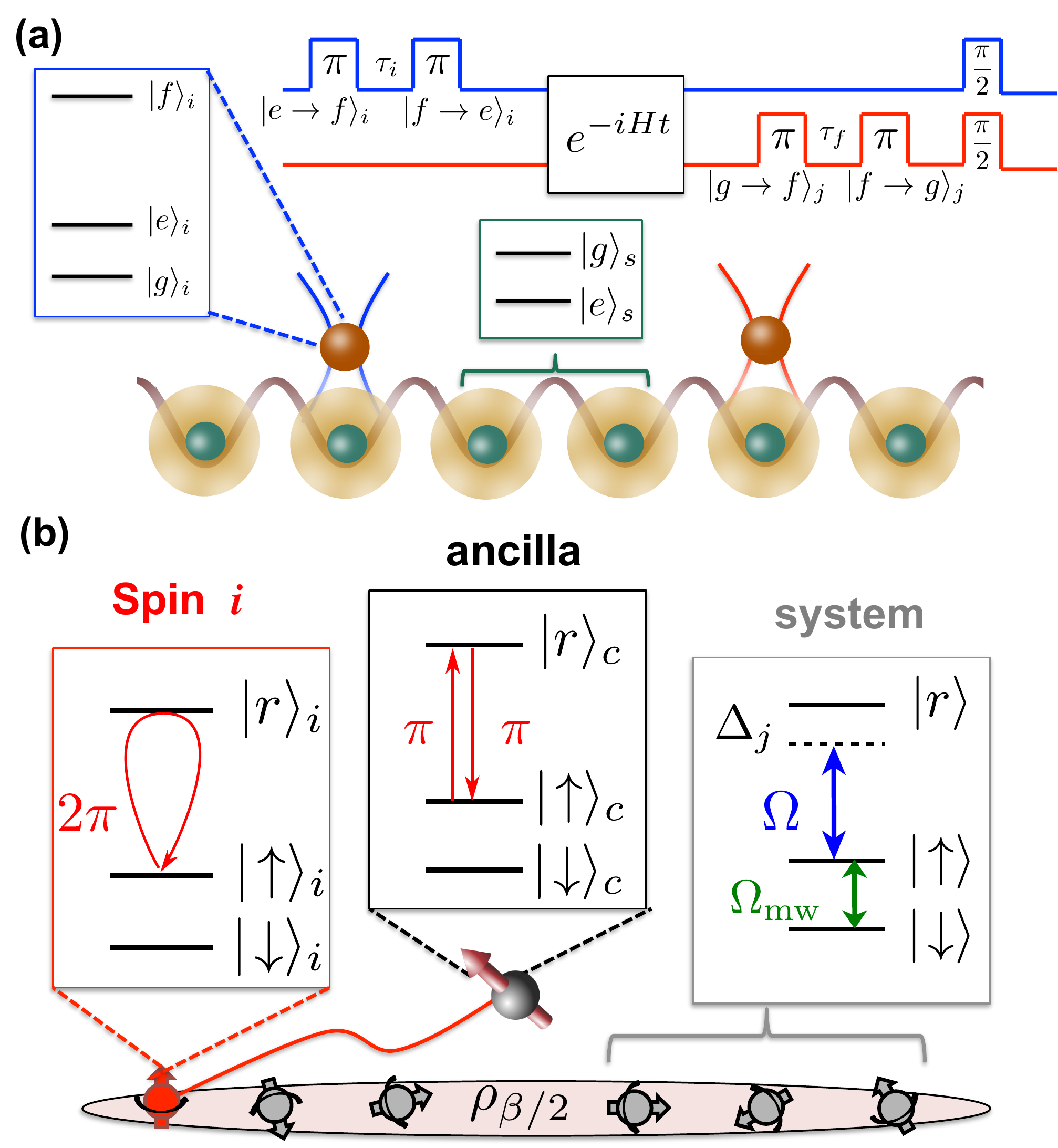}
\caption{%
\label{fig:implementation}
(a) Schematic of a one-dimensional lattice of ultracold atoms interacting with two impurities. There exists a strong interspecies Feshbach resonance between  impurity and  atom only when the impurity resides in state $|f\rangle$ . By driving the left impurity to state $|f\rangle$ before time evolution and the right impurity to $|f\rangle$ after time evolution ($\tau$ is the time needed to accumulate the Feshbach interaction), one can implement controlled unitaries $V$ and $W$  at asymmetric positions in time and space.  (b) Schematic of a 1D lattice of Rydberg atoms interacting with the ancilla via a long-range blockade-based gate. Each atom consists of two ground state hyperfine levels $\ket{\uparrow}$, $\ket{\downarrow}$ coupled to a Rydberg state $\ket{r}$ via an off-resonant laser field.}
\end{figure}

\emph{Feshbach Interactions}---Here, we envision our 1D atomic gas to be interacting with a pair of ancilla qubits (e.g. impurity atoms) at positions $i$ and $j$ (Fig.~3a). Each ancilla consists of three states $\{ |g \rangle ,  |e \rangle ,  |f \rangle \}$, with only state $|f \rangle$ exhibiting a strong inter-species Feshbach interaction with the atoms in the 1D chain. The only resource required to implement our proposed interferometric protocol is that the two ancillae are initialized in an entangled state, $|g_i g_j + e_i e_j\rangle$. In order to implement $U_{e}^\dagger = V e^{-iHt}$  and $U_{g}^\dagger =  e^{-iHt} W$, the  Feshbach interaction must be applied at different times at the two positions.
This can be accomplished via a coherent $\pi$-pulse on the ancilla at site $i$ at time $t=0$ (before Hamiltonian evolution) and on the ancilla at site $j$ at time $t$ (after time evolution), as shown in Fig.~3a.

\emph{Rydberg Blockade}---In the case where each atom within the system is  weakly dressed via a Rydberg state, $|r\rangle$, a particularly elegant approach emerges. A combination of the Rydberg-blockade  \cite{Saffman2010} and local addressability   \cite{Bakr2009,Sherson2010,Weitenberg2011} 
will then enable the application of ancilla-spin-dependent Hermitian operators (e.g. $V = \sigma_i^z$ and $W=\sigma_j^z$) at two spatially separated locations $i$ and $j$. Leveraging the long-range nature of the Rydberg interaction allows for the use of only a single ancilla atom shared between the two systems. We note that this general approach  can also be implemented in a photonic system, where long optical delay lines play the role of asymmetric timing and strong optical nonlinearities play the role of interactions.

After preparing the systems in a thermal state $\rho_{\beta/2}$ and the ancilla in $(\ket{\uparrow} + \ket{\downarrow})/\sqrt{2}$, we apply a $\pi$-pulse between the $\ket{\uparrow}_c$ and $\ket{r}_c$ states of the ancilla. Next, a $2\pi$-pulse is applied between the $\ket{\uparrow}_i$ and $\ket{r}_i$ states of atoms $i$ and $i'$ (Fig.~1 and Fig.~3b). Owing to the blockade, the $\ket{\uparrow}_i \ket{r}_c$ component of the wavefunction picks up a $\pi$ phase shift while the phases of all other components remain unchanged \cite{Jaksch2000,Lukin2001,Urban2009}. This corresponds to the application of a Hermitian operator $-2 S_i^z ( S_c^z + 1/2 )$ that implements a controlled-$V$ gate (Fig.~1b). A second $\pi$ pulse between $\ket{\uparrow}_c$ and $\ket{r}_c$ restores the state of the ancilla.  
The analogous procedure can be repeated after time-evolution to apply $W$ controlled on the ancilla qubit being in state $\ket{\downarrow}_c$, after which the resulting state $\rho(t)$ is precisely given by Eqn.~(3).

In addition to enabling our interferometric protocol, a 1D lattice of Rydberg atoms also provides a natural realization of the transverse-field Sherrington-Kirkpatrick model  [Eqn.~\eqref{tfsk}].
Experimentally, the spin degrees of freedom can be formed from two hyperfine states within each atom, wherein the transverse field $\Gamma = \Omega_{\rm mw}$ (Fig.~3b) simply corresponds to applying resonant microwave radiation between these states.
The Rydberg blockade can again be used to implement  all-to-all Ising interactions within each copy of the system. \cite{Zeiher2016}.
In particular, as shown in Fig.~3b, an off-resonant laser field with Rabi frequency $\Omega$ and detuning $\Delta_i$ couples each spin $\ket{\uparrow}$-state with a highly excited Rydberg state $\ket{r}$.
This  induces an ac Stark shift of $\ket{\uparrow}$ that depends on the spin states of surrounding atoms (e.g. due to van der Waals interactions between atoms virtually excited to \ket{r}) and  leads to a long-range interacting Hamiltonian of the form:
$ \sum_{i,j} J_{ij} \sigma_i^z \sigma_j^z$, where  $J_{ij} = \frac{\tilde{C}_6}{(r_i-r_j)^6 + \tilde{a}_B^6}$ \cite{Henkel2010}.

%
%

The coupling strengths $J_{ij}$ are determined by $\tilde{C}_6 = C_6 \Omega^4 / (\Delta_i + \Delta_j)^4$, where $C_6$ is the coefficient of the van der Waals interaction between Rydberg states $\ket{r}_i$ and $\ket{r}_j$.
In the regime where the blockade radius $\tilde{a}_B = (C_6 / |\Delta_i + \Delta_j|)^{1/6}$ is larger than the system size one obtains  all-to-all interaction between spins.
To avoid stray long-range interactions between the two copies during evolution, we assume that they are initially spatially separated, while the ancilla qubit is localized between them to allow for simultaneous coupling with each system.
Randomized Ising couplings $J_{ij}$ can be achieved by introducing a 1D speckle potential \cite{Horak1998,white2009strongly}, which leads to random detunings $\Delta_j$; if the transverse correlation length of the speckle is sufficiently large, the two copies of the system will be subject to identical randomized interactions. By introducing a second dressing laser to an attractive Rydberg state $\ket{r'}$, one can even obtain a distribution of $J_{ij}$ with zero mean \cite{Zeiher2016}.


To summarize, we have shown that concepts originating in the physics of black holes have a rich connection to quantum optical systems.
We do so through two results.  First, we introduce a general interferometric approach to measuring out-of-time-order correlation functions, which requires only local quantum control and no reverse time evolution.
Second, we show that the quantum-optics toolbox can be used to realize the TFSK model and measure $C(t)$; moreover, we demonstrate numerically that this model exhibits ``pretty fast'' scrambling.
 In particular, a phenomenological fit of $C(t)$ is consistent with a time to reach maximal scrambling proportional to $\ln N$. Because our protocol involves observables that scale with system size like $\text{Tr}(\rho^2) = e^{-S_2}$, the approach is most promising when the entropy is relatively small. Examples include mesoscopic many-body systems at low temperature, systems with an unbounded Hilbert space dimension but finite entropy, and single particle-like models that exhibit quantum chaos.

Looking forward, measurements of $C(t)$ can provide quantitative access to important questions related to thermalization and ergodicity in interacting many-body systems. For example, scrambling dynamics may give insight into the non-local dressing of Heisenberg operators under time evolution. Along these lines, it may be particularly fruitful to study integrable models, which either do not scramble or scramble only very weakly. In particular, by perturbing such models, one can probe how chaos develops and whether out-of-time-order correlation functions might exhibit evidence for a quantum analog of the celebrated Kolmogorov-Arnold-Moser theorem \cite{kolmogorov1954preservation}.
%

It is a pleasure to gratefully acknowledge the insights of and discussions with D. Stanford, C. Laumann, M. Schleier-Smith, J. McGreevy and H. Pichler. This work was supported, in part, by  NSF-DMR1507141, the Simons Foundation, the Miller Institute for Basic Research in Science, AFOSR MURI grant FA9550-14-1-0035, Harvard-MIT CUA, NSF-DMR1308435, the Gordon and Betty Moore foundation, the Humboldt foundation, and the Max Planck Institute for Quantum Optics.

\bibliography{bibFilesFabian}

\end{document}